\documentclass{article}

\usepackage{arxiv}
\usepackage{graphicx}
\usepackage{empheq,mathtools}
\usepackage{amsmath,systeme}
\graphicspath{ {./images/}}
\usepackage[utf8]{inputenc} 
\usepackage[T1]{fontenc}    
\usepackage{hyperref}       
\usepackage{url}            
\usepackage{booktabs}       
\usepackage{amsfonts}       
\usepackage{nicefrac}       
\usepackage{microtype}      
\usepackage{fancyvrb,cprotect}
\usepackage[nodayofweek,level]{datetime}

\title{Predicting the evolution Of SARS-Covid-2 in Portugal using an adapted SIR Model previously used in South Korea for the MERS outbreak}

\author{
  Pedro Teles \\
  Departamento de Física e Astronomia\\
  Faculdade de Ciências da Universidade do Porto\\
  Rua do Campo Alegre s/n, 4169-007 Porto \\
  \texttt{ppteles@fc.up.pt} \\
}

\begin{document}
\maketitle
\begin{abstract}
Since original reports in Wuhan, China, the new coronavirus covid-19 has spread very quickly worldwide, leading the World Health Organization (WHO) to declare a state of pandemic. Moreover, as of the 13th of March 2020, the WHO has announced that the European continent is now the main centre of the pandemic. Many European governments have already implemented harsh measures to attempt to contain the spread of the virus. In Portugal, there are, as of the 20th of March 2020, 785 confirmed cases. 
One of the questions many policy makers, and governments are asking themselves is how the spread is going to evolve in time. A timely idea of the amount of cases that will exist in a near future can allow governments and policy makers to act accordingly.
In this study, I applied an adapted SIR model previously used in South Korea to model the MERS outbreak, which is also caused by a coronavirus, to estimate the evolution of the curve of active cases in the case of the Portuguese situation. As some of the parameters were unknown, and the data for Portugal is still scarce, given that the outbreak started later (first case on the 2nd of March) I used Italian data (first reported case in Italy on the 31st of January) to predict them. I then construct five different scenarios for the evolution of covid-19 in Portugal, considering both the effectiveness of the mitigation measurements implemented by the government, and the self-protective measures taken by the population, as explained in the South Korean model. 
In the out of control scenario, the number of active cases could reach as much as ~40,000 people on the 5th of April (out-of-control scenario). If the self-protective and control are  taken in the same level as what was considered for the South Korean model, this number could have be reduced to about 800 cases (scenario 1). Considering that this scenario is now unrealistic. Three other scenarios were devised. In all these scenarios, the government measures had a 50\% effectiveness when compared to the measures in Korea. But in scenario 2 the transmission rate $\beta$ was effectively reduced to  50\%,  In this scenario active cases could reach circa 7,000 people. In scenario 3, the transmission rate $\beta$ was reduced to 70\% of its initial value, in which the number of cases would reach a peak of ~11,000 people. And finally in scenario 4, $\beta$ was reduced to 80\%. In this scenario, the peak would be reached at about ~13,000 cases. These 4 scenarios clearly demonstrate that self-protective measures can indeed have the much important effect of flattening the curve. In the government and the people of Portugal manage to sustain a level of control and self-protectiveness, the actual figures probably lie between the interval (~7,000-13,000) and the peak will be reached between 10th and the 20nd of April 2020.
Without control and self-protective measures, this model predicts that the figures of active cases of SARS-covid-2 would reach a staggering ~40,000 people It shows the importance of control and self-protecting measure to bring down the number of affected people by following the recommendations of the WHO and health authorities. With the appropriate measures, this number can be brought down to ~7,000-13,000 people. Hopefully that will be the case not just in Portugal, but in the rest of the World.

\end{abstract}


\section{Introduction.}
There is already abundant information on the new coronavirus and its spread worldwide \cite{1,2,3}, as well as the risks it poses to the population. The person infected with the severe acute respiratory syndrome coronavirus 2 (SARS-cov-2)\cite{4} may develop the disease called COVID-19, which manifests with mild to severe symptoms of fever, cough, shortness of breath and other signs \cite{5}. In the most severe cases, the infection can lead to the development of acute respiratory distress syndrome (ARDS) causing respiratory failure, septic shock, or multi-organ failure, and even death \cite{6}. 
There is yet a lot about this new coronavirus that is still unknown. Studies suggest that the case fatality rate of the virus is of about 3.5\% in mainland China, but the authors suggest considering the range 0.2-3.0\% \cite{7}. However, this value seems to be much higher in Italy \cite{8}, suggesting its strong dependence on demographics. In fact, the Chinese Center for Disease Control, released an analysis that showed that the case fatality rate ranged from 0\% in children aged 0-9 years old to 14.8\% in those over 80 \cite{9}, which implies that countries with a high number of elderly citizens, as is the case of Italy and Portugal, are more prone to having higher death rates. 
The quick spread of the virus in Europe has led the WHO to declare Europe as the new epicentre of the disease on the 13th of March of 2020\cite{10}. The rapid growth of the number of active cases presenting severe symptoms has saturated the health services in most countries in the continent, especially in Italy\cite{11}. It is yet unknown when the epidemic will reach an end, but governments throughout Europe have implemented harsh measures to prevent and mitigate the spread of the virus. Yet, as of today there are as many as 77,394 confirmed cases on the European continent, in 46 territories, resulting in  3,457 deaths, and rising\cite{2,3}.
In Portugal, the two first confirmed cases in the Portuguese territory were announced on the 2nd of March, which puts the country slightly behind the curve in the outbreak in the European continent. As of today, there are 642 confirmed cases, 2 deaths, and 3 recovered \cite{12}. Also, the country is on lockdown since the 13th of March 2020, and the Portuguese Parliament declared a State of Emergency on the 18th of March 2020.
One of the questions many policy makers, and governments are asking themselves is how the spread is going to evolve in time. A timely idea of the amount of cases that will exist in a near future can allow governments and policy makers to act accordingly.
In this study, I adapted a model of the SIR type used in South-Korea to model the evolution of the active cases of the MERS outbreak in the country back in 2015 \cite{13}. Although there are other models being used \cite{14,15}, this model seemed most appropriate given that the MERS and the SARS-cov-2 are related illnesses, both provoked by a coronavirus strand, and with similar symptoms, although MERS was much deadlier. In order to fit the parameters and given that there is still very little data in the case of Portugal, I used the data of Italy instead. This allowed me to fit the curve of current active cases in Portugal with a model, which I then use, by implementing the control measure parameters predicted in the model to predict the future number of cases, in five different scenarios (the out-of-control scenario, a scenario in which measures were the same as in the original model (scenario 1), a scenario where government measures are 50\% as effective as those in South Korea, and  self-protective measures reduce the transmission rate by 50\% (scenario 2), a similar scenario but in this case the transmission rate was only reduced to 70\% (scenario 3), and a fourth scenario, similar to the two previous ones but in which the transmission rate was reduced to 80\%.

\section{Methodology.}

\begin{figure}
  \centering
  \includegraphics[width=\textwidth,height=\textheight,keepaspectratio]{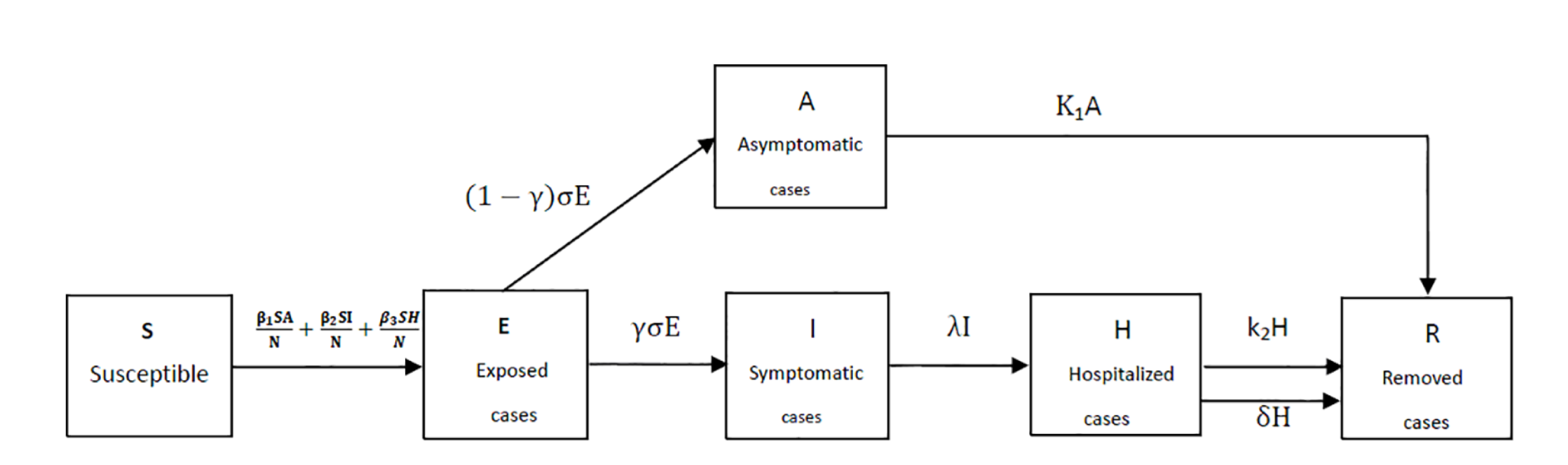}
  \caption{Flow chart of the SIR-type model used in this work, taken from \cite{13}}
  \label{fig:fig1}
\end{figure}

The model used is the model described in Xia et al\cite{13}, which can be understood by the flow diagram shown in figure \ref{fig:fig1} (which was taken from the paper, page 4). In this model, S corresponds to the number of susceptible individuals; E, the number of exposed individuals; A, the number of asymptomatic infected cases; I, the total number of mild-to-severe infected patients. H, the number of hospitalized cases; R, the number of removed cases, and finally N, the total population of Portugal.

$\beta_1$ is the transmission coefficient of the asymptomatic infected cases, $\beta_2$ is the transmission coefficient of the symptomatic infected cases (mild infected person and severe patients) to the susceptible, $\beta_3$ is the transmission coefficient of the hospitalized cases to the susceptible, $\sigma^{-1}$ is the mean incubation period, $\lambda^{-1}$ is the mean time between symptom onset to hospitalization, $k_1^{-1}$ is the mean infectious period of asymptomatic infected person for survivors, $k_2^{-1}$ is the mean duration for hospitalized cases for survivors, $\delta^{-1}$ is the mean time from hospitalization to death, $\gamma$ is the clinical outbreak rate in all the infected cases. The time unit is 1 day.

The set of differential equations can then be written as:


\begin{equation}
\systeme{
\frac{\mathrm{d} S(t)}{\mathrm{d} t} =  -\beta_1 \cdot \frac{S(t)\cdot A(t)}{N}-\beta_2 \cdot \frac{S(t)\cdot I(t)}{N}-\beta_3 \cdot \frac{S(t)\cdot H(t)}{N}, 
\frac{\mathrm{d} E(t)}{\mathrm{d} t} =   \beta_1 \cdot \frac{S(t)\cdot A(t)}{N}+\beta_2 \cdot \frac{S(t)\cdot I(t)}{N}+\beta_3 \cdot \frac{S(t)\cdot H(t)}{N}-\sigma\cdot E(t), 
\frac{\mathrm{d} A(t)}{\mathrm{d} t} =  (1-\gamma)\cdot \sigma\cdot E(t()-k_1\cdot A(t), 
\frac{\mathrm{d} I(t)}{\mathrm{d} t} = \gamma\cdot \sigma\cdot E(t()-\lambda\cdot I(t), 
\frac{\mathrm{d} H(t)}{\mathrm{d} t} =  \lambda\cdot I(t)-k_2\cdot H(t)-\delta\cdot H(t), 
\frac{\mathrm{d} R(t)}{\mathrm{d} t} =  k_1\cdot A(t)-k_2\cdot H(t)+\delta\cdot H(t)
}
\end{equation}

Given that the available data for covid-19 is still very preliminary, the model was further adapted. the only parameters that were considered “known” were the mean incubation time, which was taken as $\sigma^{-1}$ =5.1 days from the literature (here using the median as equal to the mean) \cite{16}, and the mean infectious period of asymptomatic people, where a conservative estimate of $k_1^{-1}=14$ days was used. All other parameters were taken to be “unknown”. Furthermore, and because the fitting of the model was difficult to achieve, I took $\beta_1=\beta_2=\beta_3=\beta$ as a first approximation, meaning that the mean transmission coefficient is the same independently of type and setting of the transmission. Finally, unlike Xia \textit{et al}\cite{13}, I consider that the epidemic starts with only one person infected (I0=1), being the rest of the values 0, except for $S_0$ which is simply N-1. See table \ref{table1} for a breakdown of the different parameters of this model.
 
The solving of the system of differential equations was performed using the Mathematica code \cite{17}, which also allows for the use of fittable parameters, using the function “NonLinearModelFit”\cite{18}. Given that there is still very little data for Portugal, the data for Italy was used as proxy. This seems like the most reasonable approach (instead of using data from China or another Asian country), given the cultural and health system proximity between the two countries. In order to determine the parameters $\lambda$ and $\delta$, I used the number of deaths, as reported by the Italian ministry of Health and available online [19]. For the estimation of the value of $k_2$, the number of recovered was used. All data was taken until the 19th of March 2020.
\begin{table}
\caption{The different parameters of the model described by Eq. 1}
\label{table1}
\centering
\begin{tabular}{lll}
\toprule
parameter      & Xia et al values \cite{13}  & this work  \\ \hline
$\sigma^{-1}$  & 5.2 (3$\sim$8)   (days)              & 5.1   (days)           \\
$\lambda^{-1}$ & 5(3$\sim$7)        (days)            & To be fitted     \\
$\delta^{-1}$  & 15.16(0$\sim$42)   (days)            & To be fitted     \\
$k_1^{-1}$     & 5                 (days)             &                  \\
$k_2^{-1}$     & 7              (days)                & To be fitted     \\
$\beta_1$      & 0.8756(0.853$\sim$0.9324) ($days^{-1}$)     & To be fitted     \\
$\beta_2$      & 0.7833(0.5925$\sim$0.8592) ($days^{-1}$)     & To be fitted     \\
$\beta_3$      & 0.4568(0.3839$\sim$0.6751)  ($days^{-1}$)    & To be fitted     \\
$\gamma$       & 0.0348 (0.0285$\sim$0.353) ($days^{-1}$)     & To be fitted     \\
N              & 49,520,000                     & 10,290,000       \\
$S_0$          & 49,519,960                     & 10,289,999       \\
$E_0$          & 16$\sim$32                     & 0                \\
$A_0$          & 16$\sim$36                     & 0                \\
$I_0$          & 1                              & 1                \\
$H_0$          & 0                              & 0                \\
$R_0$          & 0                              & 0                \\
\bottomrule
\end{tabular}
\end{table}

After determining the values for these parameters, they were used to fit the curve of infected cases for Portugal, where the values of  $\beta_1=\beta_2=\beta_3=\beta$ and $\gamma$ were fitted to obtain a model. This model was then used to estimate the  evolution of future cases in Portugal, according to five different scenarios:
\begin{itemize}
\item Out-of-control scenario (nothing done, the virus is free to spread (out-of-control scenario);
\item Scenario in which the government takes severe mitigating measures and the population adheres to self-protecting measures equivalent to the original model (scenario 1)
\item  Scenario where government measures are 50\% as effective as those in South Korea, and  self-protective measures reduce the transmission rate by 50\% (scenario 2)
\item Scenario where government measures are 50\% as effective as those in South Korea, and  self-protective measures reduce the transmission rate by 70\% (scenario 3)
\item Scenario where government measures are 50\% as effective as those in South Korea, and  self-protective measures reduce the transmission rate by 80\% (scenario 4)
\end{itemize}

\section{Results.}
\subsection{ Fitting of parameters $\lambda$ and $\delta$ (Italy).}

\begin{figure}
  \centering
  \includegraphics[width=\textwidth,height=\textheight,keepaspectratio]{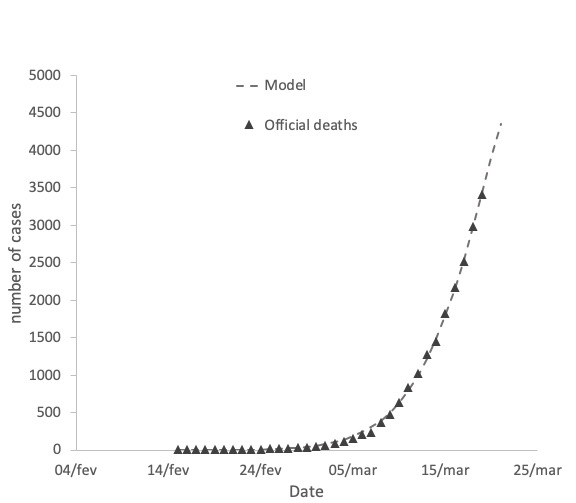}
  \caption{Graphical representation of the fitted model to the Italian government official death toll.}
  \label{fig:fig2}
\end{figure}

The number of deaths in Italy was taken from \cite{19}, and the set of differential equations described in eq. 1 fitted to obtain the best possible fit for parameters $\lambda$ and $\delta$. An $R^2$ value of 0.99255 was obtained with the fit. The rest of the parameters are presented in table \ref{table2}. Results are shown in figure \ref{fig:fig2}.
\begin{table}
\caption{Table of fitted parameters  $\lambda$ and $\delta$ as obtained using “NonLinearCurveFit” in Mathematica}
\label{table2}
\centering
\begin{tabular}{lllll}
\toprule
parameter & estimate & standard error & t-statistic & p-value               \\ \hline
$\lambda$ & 0.2548   & 0.0002         & 879         & 9.83$\times 10^{-70}$ \\
$\delta$  & 0.0168   & 0.0002         & 73          & 3.02$\times 10^{-36}$ \\
\bottomrule
\end{tabular}
\end{table}

\subsection{Fitting of parameter $k_2$ (Italy).}

Now, using the values for $\lambda$ and $\delta$ found in 3.1, parameter $k_2$ was obtained  with a new fit of Eq. 1, but now to the number of recovered cases in Italy, which again was taken from \cite{19}. The best possible fit for $k_2$ gave an $R^2$ value of 0.86178. The rest of the parameters are presented in table \ref{table3}. 

\begin{table}
\caption{Table of fitted parameter  $k_2$ as obtained using “NonLinearCurveFit” in Mathematica}
\label{table3}
\centering
\begin{tabular}{lllll}
\toprule
parameter & estimate & standard error & t-statistic & p-value               \\ \hline
$k_2$ & 0.0875   & 0.012         & 7.32         & 3.09$\times 10^{-9}$ \\
\bottomrule
\end{tabular}
\end{table}

\begin{figure}
  \centering
  \includegraphics[width=\textwidth,height=\textheight,keepaspectratio]{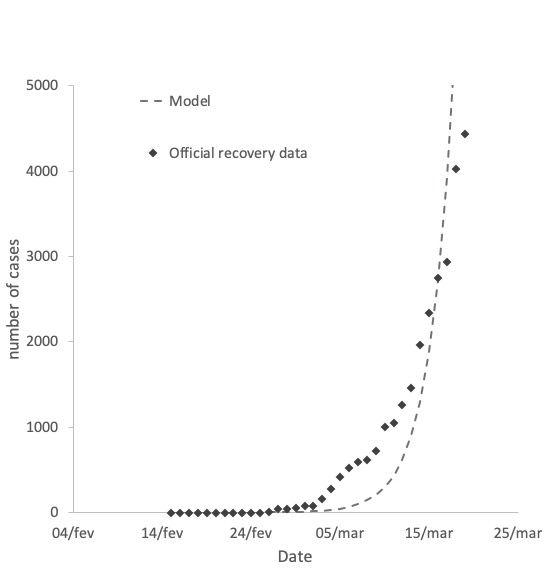}
  \caption{Graphical representation of the fitted model to the Italian government official recoveries.}
  \label{fig:fig3}
\end{figure}

As can be seen, in the case of recovered cases, the fit has not the same quality as the fit in 3.1. This is further evidenced by figure \ref{fig:fig3}, where a graphical representation of the fit is shown.

\subsection{Fitting of parameters $\beta$ and $\gamma$ (Portugal).}

After the two previous procedures, we now have all the parameters we need to fit Eq. 1 to the number of Portuguese active cases until the 19th of March, in order to determine the value of $\beta$ and $\gamma$. The number of Portuguese active cases was taken from \cite{20}, which is updated daily.

The values previously obtained of parameters $\lambda$, $\delta$, and $k_2$ were used. As explained previously, because the fitting was difficult to obtain, as a first approximation, we considered$\beta_1=\beta_2=\beta_3=\beta$. The obtained values for $\beta$ and $\gamma$ are given in table \ref{table4}.

\begin{table}
\caption{Table of fitted parameters  $\lambda$ and $\delta$ as obtained using “NonLinearCurveFit” in Mathematica}
\label{table4}
\centering
\begin{tabular}{lllll}
\toprule
parameter & estimate & standard error & t-statistic & p-value               \\ \hline
$\beta$ & 1.16   & 0.033         & 34.997         & 1.51$\times 10^{-16}$ \\
$\gamma$  & 0.0158   & 0.0034         & 4.634          & 2.74$\times 10^{-5}$ \\
\bottomrule
\end{tabular}
\end{table}

\begin{figure}
  \centering
  \includegraphics[width=\textwidth,height=\textheight,keepaspectratio]{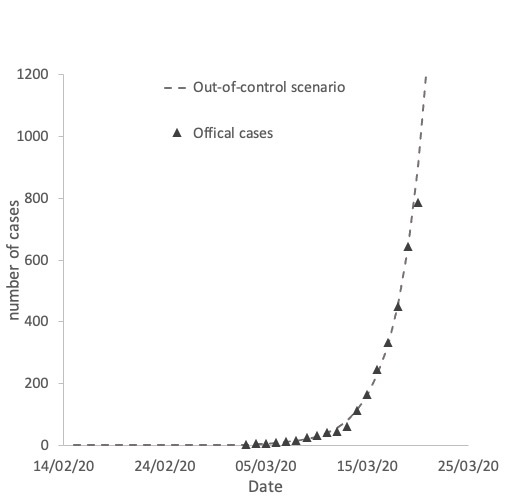}
  \caption{Graphical representation of the fitted model to the Portuguese government official active cases.}
  \label{fig:fig4}
\end{figure}

A graphical depiction of the fitted model is given in figure \ref{fig:fig4}. 

In this scenario, the out-of-control scenario, in which the corona virus would be “free” to spread within the country would lead to a peak of ~40,000 infected around the 5th of April 2020. A scenario that will very likely be avoided given the strong containment measures implemented by the government. 

\subsection{Discussion of the obtained values for $\lambda$, $\delta$, $k_2$, $\beta$, and $\gamma$}

The fitting of these parameters was difficult to obtain. There is enough systematic error and lack of statistics to make the present results very unreliable. In order to obtain the parameters for $\lambda$, $\delta$, and $k_2$, an ansatz value for $\gamma$ had to be given. In this model, $\gamma$ is originally defined as “the clinical outbreak rate in all the infected cases”. In the cases of the fits of $\lambda$, $\delta$, and $k_2$, the ersatz value used for $\gamma$ had a very strong impact in the results of the other parameters. In fact, as it comes multiplied with $\sigma$ in the transfer from compartment E(t) to compartments A(t) and S(t), I think the high variation in this value could very well be due to the huge uncertainty there is at the moment for the appropriate incubation time of this virus, which some authors claim can be of up to 27 days \cite{21}, although this number is probably an outlier. Also, a case of an incubation period of 18 days seems to have been confirmed by a recent study\cite{22}. The other parameters are somewhat less elusive. The obtained value for $\delta=0.0168$ is 50\% the value obtained by Xia \textit{et al}\cite{13} of 0.0348, s ~24\% the value obtained by Xia et al[13] of 0.0660, which shows the difference in case fatality rates between the SARS-covid-2 (7.2\% currently for Italy), and MERS (estimated at 34\%) which is ~20\%.  This seems reasonable, although the case fatality rate of the SARS-covid-2 is still difficult to ascertain. 

Finally, from the obtained values it is also possible to determine a value for the basic reproduction number $R_0$, which here I take as being simply the value $R_0=2\frac{\beta}{k_1+k_2}=14.5$ which demonstrates how strong the capacity for spreading of this particular virus is. In table \ref{table4}, the final values of the obtained parameters are given.

\begin{table}
\caption{The obtained parameters of the model described in the previous sections}
\label{table4}
\centering
\begin{tabular}{ll}
\toprule
parameter      &  this work ($days^{-1}$)   \\ \hline
$\sigma$  & 5.1(fixed)        \\
$\lambda$ & 0.2548$\pm$0.0002 \\
$\delta$  & 0.0168$\pm$0.0002 \\
$k_1$     & 0.0714 (fixed)     \\
$k_2$     & 0.0875$\pm$0.012   \\
$\beta$   &1.16$\pm$0.033      \\
$\gamma$  & 0.0158$\pm$0.034 (Portugal)\footnote{This value could go as high as 35 as an ersatz for the determination of the other values for the Italian data. Please see discussion.} \\
\bottomrule
\end{tabular}
\end{table}

\section{Five different scenarios for the evolution of SARS-covid-2 in portugal.}

The paper by Xia \textit{et al}\cite{13} further presents an extra 7 parameters to take into account control measures in time evolution of the epidemic. These parameters are meant to take into account two things:
\begin{itemize}
    \item Isolation and monitoring measures taken by the government (parameters $d_1$, $d_2$, $d_3$, and $d_4$);
    \item Self-protection measures taken by the population (parameters $l_1$, $l_2$, and $l_3$).
\end{itemize}

The way these parameters are introduced in the model are as follows:
\begin{equation}
\systeme{
\frac{\mathrm{d} S(t)}{\mathrm{d} t} =  -\beta_1 \cdot l_1 \cdot \frac{S(t)\cdot A(t)}{N}-\beta_2 \cdot l_2 \cdot \frac{S(t)\cdot I(t)}{N}-\beta_3 \cdot l_3 \cdot \frac{S(t)\cdot H(t)}{N}, 
\frac{\mathrm{d} E(t)}{\mathrm{d} t} =   \beta_1 \cdot l_1 \cdot \frac{S(t)\cdot A(t)}{N}+\beta_2 \cdot l_2 \cdot \frac{S(t)\cdot I(t)}{N}+\beta_3 \cdot l_3 \cdot \frac{S(t)\cdot H(t)}{N}-(\sigma+d_1)\cdot E(t), 
\frac{\mathrm{d} A(t)}{\mathrm{d} t} =  (1-\gamma)\cdot \sigma\cdot E(t()-(k_1+d_2)\cdot A(t), 
\frac{\mathrm{d} I(t)}{\mathrm{d} t} = \gamma\cdot \sigma\cdot E(t()-(\lambda+d_3)\cdot I(t), 
\frac{\mathrm{d} H(t)}{\mathrm{d} t} =  \lambda\cdot I(t)-k_2\cdot H(t)-(\delta+d_4)\cdot H(t), 
\frac{\mathrm{d} R(t)}{\mathrm{d} t} =  k_1\cdot A(t)-k_2\cdot H(t)+\delta\cdot H(t)+d_1\cdot E(t)+d_2\cdot A(t)+d_3\cdot I(t)+d_4\cdot H(t)
}
\end{equation}

The $l_1$, $l_2$, and $l_3$ parameters are basically multiplication factors for the transmission parameter $\beta$, so in our model we only consider one value for it as we did in the previous sections. The $d_1$, $d_2$, $d_3$, and $d_4$ parameters are summed to the parameters of spread between the infected, asymptomatic and hospitalized cases, further stressing their connection to implemented government measures. The values provided by Xia \textit{et al}\cite{13}, when applied to the Portuguese case, seem to provide a very unrealistic scenario, in which the values would immediately start to diminish. Perhaps this made sense in the case of the MERS epidemic, given that it was much more localised that the one provoked by this virus. As such, and for the lack of a better model, I devised four different scenarios in addition to the “out-of-control” scenario described in the previous section:
\begin{itemize}
\item Scenario in which the government takes severe mitigating measures and the population adheres to self-protecting measures equivalent to the original model (scenario 1)
\item  Scenario where government measures are 50\% as effective as those in South Korea, and  self-protective measures reduce the transmission rate by 50\% (scenario 2)
\item Scenario where government measures are 50\% as effective as those in South Korea, and  self-protective measures reduce the transmission rate by 70\% (scenario 3)
\item Scenario where government measures are 50\% as effective as those in South Korea, and  self-protective measures reduce the transmission rate by 80\% (scenario 4)
\end{itemize}

Obviously other values could have been used, but there was no clear indication of what these values could be, therefore I just assumed, as a first approximation, that the protective measures would manage to diminish the transmission of the contamination by 50\%. For the government related measures I used the same value as the ones from Xia \textit{et al}\cite{13}. In the second scenario I keep the 50\% reduction in the transmission rate but diminish the efforts of the government by half. In the third and fourth scenarios I use 70\% and 80\% of the transmission rate respectively.

The values for these scenarios are given in table \ref{table5}. 

\begin{table}
\caption{Control parameters of the model described by Eq. 2}
\label{table5}
\centering
\begin{tabular}{llllll}
\toprule
parameter      & Xia et al values[13]  & scenario 1& scenario 2 & scenario 3 & scenario 4 \\ \hline
$d_1$ & 0.1922  & 0.1922 & 0.0961 & 0.0961 & 0.0961      \\
$d_2$ & 0.1922  & 0.1922 & 0.0961 & 0.0961 & 0.0961      \\
$d_3$ & 0.1922  & 0.1922 & 0.0961 & 0.0961 & 0.0961      \\
$d_4$ & 0.1922  & 0.1922 & 0.0961 & 0.0961 & 0.0961      \\
$l_1$ & 0.3239  & 0.5  & 0.5 & 0.7 &0.8  \\
 $l_2$ & 0.9175  & 0.5  & 0.5 & 0.7 &0.8  \\
$l_3$ & 0.5316  & 0.5  & 0.5 & 0.7 &0.8  \\
\bottomrule
\end{tabular}
\end{table}

\begin{figure}
  \centering
  \includegraphics[width=\textwidth,height=\textheight,keepaspectratio]{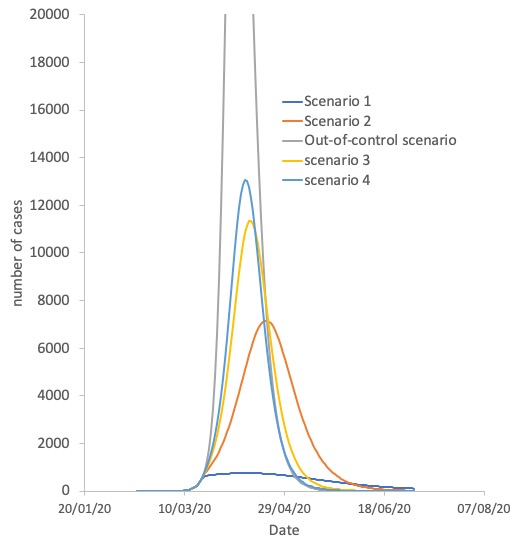}
  \caption{Graphical representation of the different scenarios considered in this study. }
  \label{fig:fig5}
\end{figure}

The day considered for the initiation of these measures is day 33 corresponding to the 18th of March 2020 which is 5 days after the government implemented the measures, so giving the virus another cycle of infection before the measures take effect. Results are shown in figure \ref{fig:fig5}.

\section{Conclusions.}

Results show that measures can be effective in “flattening the curve”, as it has come to be called this effect in recent times. In the first scenario the effectiveness of the control measures seems to be at a maximum, and a maximum of just about 800 cases would be reached. This has proven to be an unrealistic scenarion given that as of today there are 1600 active cases in the country. In scenario 2, the maximum is shifted to the 20th of April, reaching a maximum number of ~7,000 active cases in the country. In this scenario, the transmission rate has been cut to a half, but the government's measures effectiveness has been cut to a half of the value from the original model used in Korea. In scenarios 3 and 4, the measures of the government's effectiveness are kept at the same value, but the transmission rate is only cut to 70\% and 80\%, respectively. In scenario 3 the number of infected woulf reach a peak of about ~11,000 people, on the 12th of April, and finally in scenario 4, the peak would be reached at ~13,000 cases, on the 9th of April. In the out-of-control scenario, the peak would be reached at about ~40,000 people on the 5th of April.
\section{Limitations of this study and scope of application}

Given the huge uncertainties concerning these numbers, and the current lack of knowledge of the ongoing epidemic, it is very hard to make sense of its evolution, even using well-established SIR methods. The current work makes a series of assumptions which may or may not be proven correct in the future, including the incubation period of the virus, which is still not well-known. This proved to be a major source of uncertainty to fit the results, because it had a great influence in the other parameters, especially parameter $\gamma$, which seems to describe a probability of the person being exposed to the virus being symptomatic or not. Instead this parameter seems to behave in this study as an indicator of the huge uncertainty of the value for the incubation period. Finally, epidemiologists and pathologists will tell you that these parameters, unlike what is considered in this study,  not only are not constant, but dependent on time, location, culture, and other factors. It is therefore particularly hard to model the current epidemic of covid-19 at this stage.
Also, this study takes into account the particular situation for one country (Portugal), and we know that in globalised times, the likelihood that new foci of infections arise on a multi-country scale is also very high. Although currently most government have implemented serious travel restrictions, it is now well-known exactly how this parameter could affect this model, which does not take that into account. In fact, one of the major issues with the SARS-covid-2 virus is that, unlike the other SARS, it seems to be much less contained locally, as is evident by the amount of countries afflicted.
The amount of actual active cases is another clear source of uncertainty. This model takes the values of active cases and compares them with official confirmed numbers, which not only may be smaller than the actual cases, as are released at an administratively defined time, which probably does not correspond to something realistic.

The author believes that the values obtained in this study should be taken qualitatively rather than quantitatively. One can think of the scenarios as defining two limits of a likelihood interval, and as such, the actual figure of maximum number of active cases in Portugal will likely be in the interval ~7,000-13,000 between the 9th-20th of April, if the government and the people of Portugal manage to keep implementing self-protective and control measures. This can perhaps serve to advise the competent authorities in taking the appropriate measures. It also serves to show the importance of control and self-protecting measures. It is possible to bring down the number of affected people by following the recommendations of the WHO and health authorities. Without them, this model predicts that the figures of active cases would reach a staggering ~40,000 people. It also shows that with the appropriate measures this can be indeed brought down, thus flattening the curve. Hopefully that will be the case not just in Portugal, but in the rest of the World.


\end{document}